\documentstyle[pre,aps,epsfig,floats]{revtex}

\begin{document}
\draft

\title{Effect of Size Polydispersity on Melting of Charged Colloidal Systems\footnote{ Chinese. Phys. Lett. 20, 1626(2003)}}

\author{Yong Chen\thanks{Email address: ychen@lzu.edu.cn.}}

\address{Institute of Theoretical Physics, Lanzhou University, Lazhou 730000}

\date{\today}

\maketitle

\begin{abstract}
We introduce simple prescriptions of the Yukawa potential to describe the effect of size polydispersity and macroion shielding effect in charged colloidal systems. The solid-liquid phase boundaries were presented with the Lindemann criterion based on molecular dynamics simulations. Compared with the Robbins-Kremer-Grest simulation results, a deviation of melting line is observed at small $\lambda$, which means large macroion screening length. This deviation of phase boundary is qualitatively consistent with the simulation result of the nonlinear Poisson-Boltzmann equation with full many-body interactions. It is found that this deviation of the solid-liquid phase behaviour is sensitive to the screening parameter.
\end{abstract}


Charged colloidal suspensions have received much attention in recent years because of their abundant structural and rheological behaviour. As a classical complex many-body system, they are also an impressive model system since it is easy to be verified by various optical techniques.$^{\cite {Loewen 1994}}$ The classical explicit expression of effective pair interaction of one-component model of colloidal systems is the Dejaguin-Landu-Verwey-Overbeek (DLVO) potential, which means the interaction between an isolated pair of spherical macroions with finite size. The DLVO potential consists of an electrostatic and the van der Waals part.$^{\cite {Verwey 1948}}$ In general, it is able to manage with Yukawa or screened Coulomb form which only retains the electrostatic part. Furthermore, the phase diagram of Yukawa systems was presented by computer simulations.$^{\cite {Kremer 1986,Robbins 1988}}$ The phase transition line from order (bcc or fcc) to disorder (liquid) is approximately given by a straight line, $\bar{T} _M=0.00246+0.000274\cdot \lambda $, where $\bar{T}_M$ is the rescaled melting temperature and $\lambda $ is a measure of verse rescaled screening parameter.

However, in actual colloidal suspensions, the particles are not identical in size, or they exist size polydispersity. In recent experiments Yamanaka {\it et al.} studied the reentrant crystalline-liquid transition induced by surface charge.$^{\cite{Yamanaka 1998,Yamanaka 1999,Yamanaka 2002}}$ A visible discrepancy of phase boundary between size polydispersities $2\%$ and $4\%$ of latex is observed. On the other hand, Brunner {\it et al.} experimentally investigated the effective interparticle potential in two-dimensional charge-stabilized colloidal system at various particle densities.$^{\cite{Brunner 2002,Klein 2002,Russ 2002}}$ It is found that the effective pair colloid-colloid potential is in agreement with the Yukawa form of the DLVO theory only at low density, but not at high density. A suggestion, the so-called macroion shielding, has been put forward to explain the density-dependent pair interaction.

It is well-known that the macroscopic properties of charged colloidal suspension are determined by microscopic interaction among the constituents of this system. The main purpose of the present work is to investigate the influence of macroscopical solid-liquid phase behaviour caused by size polydispersity and macroion shielding. Through picturing these effects as simple prescriptions of Yukawa pair interaction, we present the melting lines based on molecular dynamics (MD) simulation. A very similar deviation of the solid-liquid phase boundary is observed at large screening length of macroion. This indicates that the screening parameter plays an important role in the macroscopic properties of charge-stabilized colloidal suspensions.

In the DLVO description of charged colloidal system of $N$ macroions with number density $\rho$, the resulting effective pair potential between macroions is of the Yukawa shape with length scale $a=\rho ^{-1/3}$,$^{\cite{Verwey 1948}}$ \begin{equation}
U\left( r\right) =A\frac{\exp (-\lambda r/a)}{r/a},  \label{eq-1}
\end{equation}
where $r$ is the center-center separation between two particles; $\lambda =\kappa a$ with $\kappa $ being the Debye-H\"{u}kel screening parameter;  and $ A=\left( Z^{*}e\right) ^2/\left( \epsilon a\right) $ is the prefactor to describe the interaction strength with $Z^{*}$ being the charge of particle, $e$ is the charge of an electron in electrostatic units, and $ \epsilon $ is the background solvent dielectric constant. It is noted that in this description of prefactor, we have incorporated the size correction factor directly into the effective charge. Clearly, the Yukawa picture of colloidal systems only depends on the rescaled separation between particles $r/a$ and the parameters $A$ and $\lambda $.

In the course of our work on melting lines, we follow the simulation scheme of Robbins{\it et al.}$^{\cite{Kremer 1986,Robbins 1988}}$ It is worthwhile to normalize the parameters with typical phonon energy. As a general rule, the rescaled temperature is defined by

\begin{equation}
\bar{T}=K_BT/\left( ma^2\omega _E^2\right),  \label{eq-2}
\end{equation}
where $K_BT$ is the thermal energy, $m$ is the mass of particle, and $\omega _E$ is the Einstein frequency. Moreover, from the calculation of the total energy per particle at zero temperature (all particles are posited at the lattice sites), it is easy to characterize the prefactor of pair potential by Einstein frequency $\omega _E$.

In the sense of intuitive physics, the direct solution of studying the effect of size polydispersity is to deduce the fluctuation distribution of prefactor in the Yukawa potential from the size distribution of particles. However, it is a puzzle how to establish the dependence correlation between the distribution of effective surface charge and the size distribution. It is concluded that the different presuppositions, constant charge density and constant electrical potential on the surface of particle would lead to conflicting results. On the assumption of constant surface electrical potential, one could bring on the DLVO Yukawa pair potential, whether by analytic deduction from the mean spherical approximation of primitive model or by numerical simulation from complete Poisson-Boltzmann equation.$^{\cite{Noyola 1980,Hoskin 1956,McCartney 1969,Bell 1970}}$ On the other hand, for the assumption of constant surface charge density, it is proven from the hypernetted-chain approximation that the colloid-colloid interaction is not a simple Yukawa shape with neglect of van der Waals forces and it is possible to perform the attractive interaction at small separation under some conditions.$^{\cite{Patey 1980}}$

In the viewpoint of theoretic works, with discarding the exact distribution function of charge numbers of particles caused by size polydispersity in charged colloidal systems, we have just introduced a simple prescription into the Yukawa potential, i.e.,

\begin{eqnarray}
U & = & A^{\prime }\frac{\exp (-\lambda r/a)}{r/a} \nonumber\\
A^{\prime } & = & A\left( \gamma +1\right), \label{eq-3}
\end{eqnarray}
where $A$ is the prefactor in Eq. (\ref{eq-1}) with its value determined from Table 1 in Ref. {\cite{Robbins 1988}} and $\gamma$ is a Gaussian random number with mean value zero and variance $\sigma$. In Eq. (\ref{eq-3}), the effect of size polydispersity is approached by a Gaussian fluctuation of prefactor. It is not a accurate detailed description of size effect. Even so, this coarse-grained prescription is enough to extract very useful information about the effect of size polydispersity on the solid-liquid phase behaviour of charged colloidal suspensions.

To simulate the solid-liquid phase boundary, in strictly speaking, one should calculate the free energies for solid and liquid states. In the past decades, there are many computer simulations for quantitative results and relative theoretic techniques about criterion of this phase transition, melting or freezing. All of these progresses are contributed by the advanced experiments and well-characterized colloidal systems. In this work, we are only interested in the solid-liquid phase transition, or melting line. Normally, one should be considered the Lindemann criterion or the Hansen-Verlet rule in the calculation. The Lindemann criterion states that the ratio of the root-mean-square (rms) displacement $\sqrt{\left\langle \delta u^{2}\right\rangle}$ and the characteristic interparticle distance or $a$ of the solid at the $\bar T_{M}$, the melting temperature, has a universal value.$^{\cite{Lindemann 1910}}$ The Hansen-Verlet rule means that the first maximum of liquid structure factor has an amplitude of about $2.85$ along the phase transition line.$^{\cite{Hansen 1969}}$ Both criteria would lead to the identical melting temperature in this simulation scheme. However, the rms displacement is easier to be calculated and gives the more precise values of $\bar T_{M}$.$^{\cite {Robbins 1988}}$ Consequently, the rms displacement is calculated to achieve the melting line according to the Lindemann criterion, $\sqrt{\left\langle \delta u^{2}\right\rangle} /a=0.19$, in all of our simulations.

A very important simulation technique, constant temperature MD method, is employed to realize calculation of the rms displacement at expectant temperature. The so-called stochastic method that the experiment system is coupled with a head bath is used in our simulations. For every particle with mass $m$ interacted by a total Yukawa potential $U_t$ from other particles, one can consider the equations of motion for Cartesian coordinate,$^{\cite {Robbins 1988,Nose 1991,Schneider 1978}}$

\begin{equation}
m\frac{d^2{\bf r}_i}{dt^2}=-\frac{\partial U_t }{\partial {\bf r}_i} -\eta \frac{d{\bf r}_i}{dt}+{\bf R}_i\left( t\right),
\label{eq-4}
\end{equation}
where a friction force with coefficient $\eta$ and a random force ${\bf R_i\left(t\right)}$ are contacted. The amplitude of random force is governed by the fluctuation dissipation theorem, $\left\langle {\bf R}_i\left( t_1\right) {\bf R}_j\left(t_2\right) \right\rangle =2K_BT\eta \delta _{ij}\delta \left(t_1-t_2\right)$, where $T$ denotes the temperature of heat bath. This solution of constant MD simulation is easy not only to put into practice, but also to obtain excellent performance. In our simulations, the average relative error between the real stable temperature and the expectant temperature is less than $1\%$.

All of our simulations was carried out with $N=432$ and $500$ particles, for initial configurations of bcc and fcc crystals in cubic box with periodic boundary, respectively. The time step in simulations is $\Delta t=0.01\tau _E$ where $\tau _E=2\pi /\omega _E$ is from the definition of energy scale. The time to reach equilibrium status for experimental systems is not longer than $10000$ time steps, and the all runs is $100000$ time steps.

\begin{figure}[h]
\begin{center}
\epsfig{figure=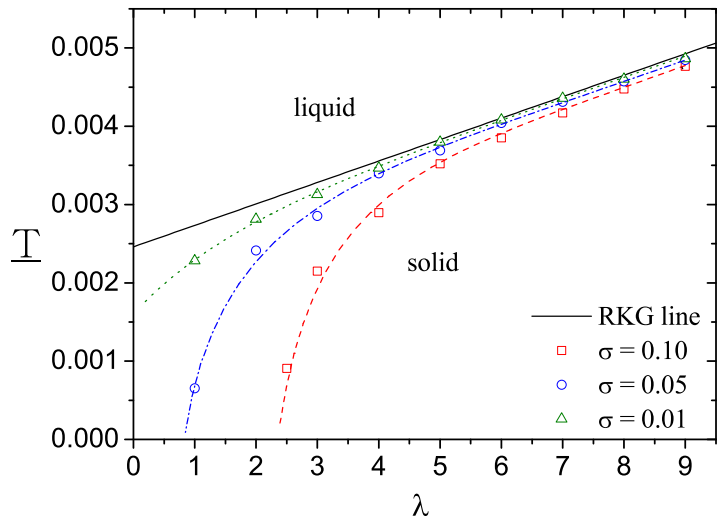,width=11cm}
\caption{Melting lines of Yukawa systems with various fluctuations of prefactor. The solid line is the RKG melting line without fluctuation of prefactor. The length of cutoff for pair interaction $r_c=3.07a$ for all phase boundaries. The squares, circles, and triangles indicate the values of $\bar {T_M}$ where the rms displacement reached $0.19$ of $a$ from our MD results.}
\label{Fig1}
\end{center}
\end{figure}

In Fig.\ref{Fig1}, we presented the melting lines of charged colloidal system with our simple prescription Eq. (\ref{eq-3}) for varied distributions. The open squares, circles and triangles are the melting points from our MD simulations according to the Lingdemann criterion for the variance of fluctuation $\sigma=0.10$, $0.05$ and $0.01$, respectively. The solid line is the RKG melting line and the others are the fitting curves. Obviously, at small $\lambda$, or large Debye screening length, the deviation of solid-liquid phase boundary is observed. Moreover, the larger $\sigma$ from the size polydispersity becomes, the larger deviation of melting line becomes. For the larger values of $\lambda$, however, it is just a very little shift of phase boundary compared with the RKG line.

In a general way, the interaction between particles is characterized with the new length scale $a$, $U_{a}=U_{0}$exp$(-\lambda)$.$^{\cite{Robbins 1988}}$ In the experiment on phase diagram of charged colloidal suspensions,$^{\cite {Monovoukas 1989}}$ it proves very good agreement between the experimental results and the RKG transition lines after using renormalized charge. For example, from Fig. 8(b) in Ref. \cite{Monovoukas 1989}, $K_{B}T/U_{a}$ is about $0.15$ for $\lambda=5$, but from the RKG prediction without renormalization, this value is $0.1954$. In our simulation result as shown in Fig.\ref {Fig1}, $K_{B}T/U_{a}$ is $0.1937$, $0.1883$, and $0.1637$ for $ \sigma=0.01$, $0.05$, and $0.10$, respectively.

A qualitative rational explanation follows from the deviation in Fig.\ref{Fig1} at small $\lambda$. The deviation of the solid-liquid phase boundary arises from the correlated fluctuations of the ion clouds around macroion. In a general case, this effect is not strong enough to influence the phase transition especially in monovalent electrolytes. However, this correlated effect has been enhanced by size polydispersity. As a result, the noticeable deviation is found at large polydispersity, such as $\sigma = 0.05, 0.10$ in Fig.\ref {Fig1}. For $\sigma =0.01$, the melting line is very close to the RKG line since the deviation at small $\lambda$ can be accepted after considering the simulation error. On the part of large $\lambda$, the deviation has reduced to a very small shift because the ionic distribution about the central macroion decreases dramatically with distance and the polydispersity scarcely affects the correlated effect. As to the shift of phase boundary in Ref.\cite{Monovoukas 1989}, is is caused by the nonlinear effect in expansion of Possion equation. Anyway, in the physical nature, the all shifts of phase boundary is from the microscopic interaction although one root in the correlated fluctuation of ionic cloud and another comes from the nonlinear effect in expansion of Poisson equation. The detailed microscopic origins will be studied in the future works.

All the above calculations are completed with the cutoff of pair interaction $r_c=3.07a$, as the same as for the RKG line, since the longer range contribution to the energy is close to an independent structure.$^{\cite{Robbins 1988}}$ Acting up to the physical instinct, someone may image that the length of cutoff is longer, and the simulation result is better. However, one should note that the Yukawa potential is only interaction between two particles with finite size, and is deduced with the presupposition of dilute limit of particle density. In fact, the three-body interaction and more order effects are obliterated in the calculation with this form of pair interaction. Based on the recent experimental results about effective pair interactions at different densities,$^{\cite{Brunner 2002,Klein 2002}}$ we have proposed a simple prescription to the Yukawa potential,$^{\cite{Dobnikar 031,Dobnikar 032,Dobnikar 033}}$

\begin{eqnarray}
U(r) & = & A\frac{\exp (-\lambda r/a)}{r/a} \quad r\leq r_c \nonumber\\
 &=&0  \quad r>r_c  \label{eq-5}
\end{eqnarray}
with prefactor $A$, and a density-dependent cutoff $r_c\propto a=\rho^{-1/3}$ to include the macroion shielding effect.

\begin{figure}[h]
\begin{center}
\epsfig{figure=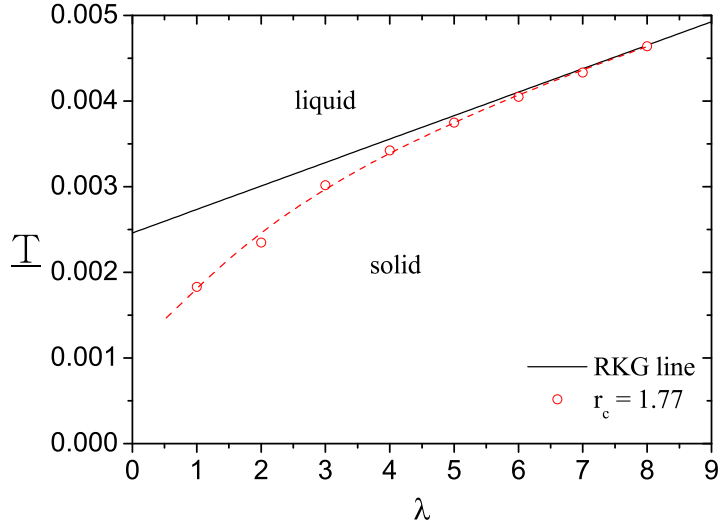,width=11cm}
\caption{Melting line of Yukawa systems with length of cutoff $r_{c}=1.77a$. There is no effect of size polydispersity added to experiment systems. The open circles are the values of $\bar {T_M}$ from the Lindemann criterion in our MD simulations. }
\label{Fig2}
\end{center}
\end{figure}

Based on this simple modified pair potential, the solid-liquid phase boundary for $r_c=1.77a$ is plotted in Fig.\ref{Fig2} as a dashed line and the open circles are the rms displacement calculated from our MD simulations with the Lindermann criterion. The solid line, $\bar T_M=0.00246+0.000274\lambda$, is the RKG melting line produced for $r_c=3.07a$.$^{\cite{Robbins 1988}}$ It is easy to observe that a visible discrepancy is observed for small $\lambda$, just like that in Fig.\ref {Fig1}. However, for a larger value of $\lambda$, the melting points are almost identical to the RKG results and there are no shifts for all $\lambda$. As interpretation mentioned above, the density-dependent truncation of pair interaction is resulted from the many-body effect in concentrated charge colloidal suspensions. It follows qualitatively that the macroscopical phase behaviour is only influenced noticeably for small $\lambda$ (or large Debye screening length) from Fig.\ref{Fig2}. This result is consistent with the conclusion from simulations with a continuous Poisson-Boltzmann description which includes fully many-body effects in colloidal systems.$^{\cite {Dobnikar 031,Dobnikar 032,Dobnikar 033}}$

To compare the above two simulation results, Fig.\ref{Fig1} and Fig.\ref{Fig2}, we have observed a similar deviation of solid-liquid phase behaviour in the area of large screening length. However, in Fig.\ref{Fig1}, this discrepancy is from the fluctuation of interactions between particles, which is enhanced by the size polydispersity of colloidal particles. Figure \ref{Fig2} shows a result from the macroion shielding effect which is due to the many-body effect at high density.

It is perhaps worthwhile to comment many-body attraction. It is well known that an attractive component in pair interaction can appear due to the correlated effect. As a straightforward result of strong correlation effect from large size polydispersity, the pair potential should include an attraction part. In our simulation for Fig.\ref{Fig1}, avoiding the difficulty to obtain the exact uniform effective pair interaction, the correlated fluctuation of ionic cloud is directly and approximately reflected by Gaussian distribution of the prefactor in the Yukawa pair potential. On the other hand, the charge-like attraction is obtained and it becomes visible with the increasing particle density.$^{\cite{Brunner 2002,Klein 2002}}$ It is noted that the effective part in pair potential is similar to the same part of the Yukawa potential after considering the density-dependent truncation of interaction length to include many-body effect. In a sense, we have conjectured that the deviation of macroscopic phase behaviour corresponds to appearance of many-body attraction especially at low salt concentration for high particle density or large size polydispersity.

In conclusion, considering size polydispersity and macroion shielding effect as simple prescriptions of the Yukawa potential between colloidal particles, we have performed MD simulation results of solid-liquid phase behaviour. A visible deviation of phase boundary is observed at small $\lambda$, or large screening length of macroions. It is found that this discrepancy is sensitive to the Debye screening length. Again, it is emphasized that the screening parameter plays a major role in the interpretation of macroscopic properties of charged colloidal systems.

\bigskip

We gratefully acknowledge Jupei Yamanaka for useful and stimulating communications.

\end{document}